\documentclass{aa}  

\usepackage{graphicx}
\usepackage{txfonts}
\usepackage{xcolor}
\usepackage{hyperref}
\hypersetup{
    colorlinks=true,
    citecolor=violet,
    linkcolor=teal
}

\usepackage{natbib}
\bibpunct{(}{)}{;}{a}{}{,} % to follow the A&A style
\newcommand{\msun}{$M_\odot$\xspace}

\begin{document} 

   \title{The first phase of mass transfer in low-mass binaries: Neither stable nor a common envelope}

   \author{Gijs Nelemans
          \inst{1,2,3}
          \and
          Holly Preece
          \inst{1}
          \and
          Karel Temmink
          \inst{1}
          \and
          James Munday
          \inst{4}
          \and Onno Pols
          \inst{1}
          }

   \institute{Department of Astrophysics/IMAPP, Radboud University,
             P.O. Box 9010, NL-6500 GL Nijmegen, The Netherlands
             \email{nelemans@astro.ru.nl}
        \and
          Institute of Astronomy, KU Leuven, Celestijnenlaan 200D, B-3001 Leuven, Belgium
        \and
         SRON, Netherlands Institute for Space Research, Niels Bohrweg 4, 2333 CA Leiden, The Netherlands   
        \and
        Department of Physics, University of Warwick, Gibbet Hill Road, Coventry, CV4 7AL, United Kingdom
        }

   \date{Received April 11, 2025; accepted June 18, 2025}

% \abstract{}{}{}{}{} 
% 5 {} token are mandatory
 
  \abstract
   {The masses of the white dwarfs in a binary carry information about previous mass-transfer phases. The core mass -- radius relation of low-mass giants gives the size of the progenitor of a helium white dwarf at the moment it last filled its Roche lobe. Previously, we used this information for a few observed systems to propose a new mass-transfer type based on an angular momentum balance.}
   {Our aim is to investigate if stable mass transfer instead of the angular-momentum prescription is consistent with the observed double-helium white-dwarf masses.}
   {We reconstructed the progenitor evolution of observed double-helium white dwarfs using the core mass -- radius relation and evaluated if the periods at the start of the second phases of mass transfer are consistent with the outcome of stable mass transfer. More generally, we calculated the mass distribution of double-helium white dwarfs for three different progenitor scenarios: double common envelope (with parameter $\alpha \lambda$), angular-momentum prescription (with parameter $\gamma$), and stable mass transfer.}
   {We find that the observed systems are generally not consistent with stable mass transfer. Stable mass transfer leads to a  tight correlation between the two white dwarf masses in a binary that is not consistent with the observed mass distribution. Double-common-envelope evolution is a particularly poor fit to the observations. The angular-momentum prescription can populate the observed mass distribution, but not perfectly.}
   {We conclude that the first phase of mass transfer initiated on the red giant branch in low-mass systems does not generally proceed as stable mass transfer nor as a common envelope, and thus it is poorly understood. This may be related to the fact that for many observed binaries that have finished the first phase of mass transfer the orbit is eccentric, which is an unexpected outcome of mass transfer. }


   \maketitle
%
%-------------------------------------------------------------------

\section{Introduction}

Most stars in the universe have low masses, and all of these low-mass stars ($\sim$0.8-2\msun) follow a similar evolution \citep[e.g.][]{1971A&A....13..367R,1971ARA&A...9..183P}: after the main sequence the helium core contracts, becomes degenerate, and grows in mass via hydrogen-shell burning. At the same time, the envelope of the star expands, and the star ascends the giant branch until the core experiences the helium flash around a mass of 0.47\msun. The expansion of these stars happens for a significant part when they have deep convective envelopes. 

For a long time, it was believed that the first mass-transfer phase for binary systems with such stars (where the mass ratio $q = M_\text{donor}/M_\text{accretor} > 1$) would lead to a runaway mass transfer and the formation of a common envelope, unless their period is very short \citep[e.g.][]{1976IAUS...73...75P,1988SvAL...14..265T}. The common envelope is expected to lead to severe orbital shrinkage, inspired by the observation that many close binaries with white-dwarf (WD) components have orbits that must have shrunk a lot \citep{1976IAUS...73...75P}. However, many observed post-mass-transfer binary systems containing low-mass stars, including barium stars and blue stragglers in clusters, have orbital periods that are relatively wide (and often eccentric) that do not fit with this picture \citep[see][and references therein]{2019MmSAI..90..395J}. 

Strong orbital shrinkage in the first phase of mass transfer also leads to problems with double WDs that would go through a common-envelope phase twice. The second common envelope would then necessarily happen in a tighter binary than the first, and the strict core mass -- radius relation of giants with degenerate cores would then produce double WDs where the younger (second) WD is significantly lower in mass than the older (first formed) one \citep[e.g.][]{1995MNRAS.272..800H,1997ApJ...475..291I,1998MNRAS.296.1019H}, contrary to many observed systems \citep{1999ASPC..169..275M,2000A&A...360.1011N}. 

Two solutions to this problem have been put forward. The first one suggests that the first common envelope is qualitatively different from the second, due to the larger mass ratio. This implies a larger angular momentum in the binary system and the phase between the onset of Roche-lobe overflow and the actual engulfment would take longer (if the engulfment happens at all), leading to wider orbits after the interaction. A formalism was proposed to calculate the outcome based on angular momentum, assuming that the mass lost from the binary takes away angular momentum per unit mass; that is, a constant $\gamma$ times the average angular momentum per unit mass. This '$\gamma$' formalism was found to explain the observations well\footnote{This is somewhat surprising, since (as people have brought forward as criticism of this approach) the strong sensitivity of the orbital change to $\gamma$ could artificially lead to a narrow range of acceptable $\gamma$-values from the observations.  However, then using $\gamma$ in forward modelling would lead to erroneous results, contrary to our findings \citep[e.g.][]{2001A&A...365..491N,2012A&A...546A..70T}.}\citep{2000A&A...360.1011N} for values of $\gamma \approx 1.5-1.75$.

A second suggestion is that the mass transfer is in fact stable \citep[e.g.][]{2008ASSL..352..233W,2012ApJ...744...12W,2023A&A...669A..82L,2024ApJ...977...24Z}. \citet{2008ASSL..352..233W} postulated the formation of all double WDs from the small initial range of binaries that interact before having a deep convective envelope. That assumption cannot explain the relatively large number of systems. \citet{2012ApJ...744...12W} find models where mass transfer is stable, 
despite the convective envelope of the donor. More recently, a number of studies also find that the reaction of the donor to mass loss is much more favourable for stability than was previously thought \citep[e.g.][]{2020ApJ...899..132G,2023A&A...669A..45T}. In addition, some objects are found that fall near the period -- WD mass relation expected after stable mass transfer \citep{2023MNRAS.518.4579P}. \citet{2024ApJ...977...24Z} studied the formation of double-helium WDs through stable mass transfer plus common-envelope evolution and used that to constrain the common-envelope phase. \citet{2023A&A...669A..82L} investigated the effect of the assumption that the first phase of mass transfer was stable on the population of double WDs and find that it reduces the number of double WD.  Double WDs are more widely relevant, for example for type Ia supernovae \citep[e.g.][]{2014ARA&A..52..107M,2023RAA....23h2001L,2024arXiv241201766R} and as gravitational wave sources for the LISA mission \citep[e.g.][]{2001A&A...365..491N,2010ApJ...717.1006R,2010Ap&SS.329..297L,2010A&A...521A..85Y,2017MNRAS.470.1894K,2019MNRAS.490.5888L,2023LRR....26....2A,2024MNRAS.534.1707T}, so it is important to understand their formation. 

In this work, we investigated the possibility of observationally testing the hypothesis that the first phase of mass transfer was generally stable. Many of the binaries that we observe directly after this phase are difficult to interpret due to very strong selection effects, as they are found through their peculiar abundances or in clusters, and the non-evolved star easily outshines the WD. We therefore used observations of systems that have evolved further, via a second mass-transfer phase to become double WD or subdwarf B (sdB) + WD systems. The first phase of mass transfer produces a WD that we observed as the older WD, while the second mass-transfer phase produces the observed younger WD or the sdB. As shown in previous work \citep[e.g.][]{2000A&A...360.1011N,2011MNRAS.411.2277D}, the period before the second mass-transfer phase, and thus after the first one, can be reconstructed from the observations \citep[also giving information about the common envelope  in the second phase of mass transfer, e.g.][]{2011MNRAS.411.2277D,2023MNRAS.518.3966S}. 

 In Sect.~\ref{sec:methods}, we  detail the methods we used. In Sect.~\ref{sec:observations}, we describe the data from which we derived our results; these are presented in Sect.~\ref{sec:results}. We discuss our results and conclude in Sect.~\ref{sec:conclusion}.

 \section{Method}\label{sec:methods}

The formation of a close pair of WDs generally proceeds as follows: the most massive star (primary, indicated with subscript 1 below) evolves to become a red giant and at some point in its evolution fills its Roche lobe and begins mass transfer. After the mass transfer, the remaining core of the most massive star becomes a WD (WD1). The initially less massive star (secondary, indicated as star 2 below) evolves later and again starts mass transfer when it fills its Roche lobe; this typically happens on the giant branch, and then a second WD (WD2) is formed. Both WDs might first be sdB stars for a short period, if their mass is close to 0.47\msun.

For low-mass stars, $M \la 2$\msun, the evolution after the main sequence quickly reaches the red giant branch, where these stars develop degenerate helium cores. Due to the tight core mass -- luminosity relation of such giants and the fact that they have deep convective envelopes and thus reside on the Hayashi track, there also is a tight core mass -- radius relation for such giants. There are a number of different studies that determine the core mass -- radius relation (or, alternatively, the period -- core mass relation for giants filling their Roche lobe). Based on these studies, we constructed the two curves we use in this paper (see Appendix~\ref{sec:Appendix_fits} for more details): one for approximately solar metallicity ($Z\approx0.02$) and one for intermediate metallicity ($Z\approx0.0045$); 
\begin{equation}\label{eq:R_Mc}
    R (M_c) = R_* \frac{M_c^6}{(1 + 45 M_c^5)} R_\odot
,\end{equation}
with $R_* = 3 \times 10^4$ for solar metallicity and $2.4 \times 10^4 $ for intermediate metallicity. These bracket the bulk of the star formation in the Milky Way (see Appendix~\ref{sec:Appendix_fits}). At a very low core mass and short period, the core mass -- period relation may not always hold \citep{2017MNRAS.467.1874C,Chen_Tauris_Chen_Han_2022}, and in general this relation is somewhat model dependent (see Appendix~\ref{sec:Appendix_fits}).

Because the core mass -- radius relation is very well defined for giants with degenerate helium cores, we restricted ourselves to systems for which Roche-lobe overflow in this first mass-transfer phase leads to the formation of a helium WD. These have  masses in the range between the minimum core mass at the base of the red giant branch ($\ga 0.15$\msun) and the maximum core mass, at which point the core undergoes the helium flash ($\sim0.47$\msun). We investigated different scenarios of how this first phase of mass transfer may proceed.

A second phase of mass transfer leads to the formation of a double-WD system. Because of the large mass ratio at the onset of this second mass transfer (donor with $M > 1$\msun and accretor with $M_{WD} < 0.5$\msun) as well as the short periods of the current orbits, we assume that the second phase of mass transfer must have been a common-envelope phase with significant shrinkage of the orbit. We also assume that this happened on such a short timescale that the core mass of the giant at the start of the mass transfer is equal to the mass of the second formed WD ($M_{WD2}$).

%--------------------------------------------------------------------

\subsection{Reconstructing the evolution using the observed masses}

We can reconstruct the evolution of the binary based only on the masses by using the core mass -- radius relation, together with the fact that the stars that follow this type of evolution have masses in a relatively small range from about 1\msun to $\sim$2\msun. Below 1\msun, the stars do not evolve in the age of the Universe, above $\sim$2\msun, the stars do not form degenerate cores on the red giant branch. A very similar approach was taken by \citet{2024ApJ...977...24Z}. 

Let us denote the initial masses of the two stars as $M_1$ and $M_2$ and the initial orbital period as $P_i$ and initial separation $a_i$. We use $m$ as subscript for the properties after the first mass transfer, but before the second mass transfer. The masses of the WDs formed from the primary and secondary are indicated with $M_{WD1}$ and $M_{WD2}$. The latter is observed as the younger, brighter object and thus in observational work is typically referred to as $M_1$. We neglected initial eccentricity, as the orbit is expected to circularise before mass transfer starts. The primary fills its Roche lobe at the moment that 
\begin{equation}
    R_1 = R_{L,1} = r_{L}(q) \, a_i
,\end{equation}
with $q = M_1/M_2$ and $r_L(q)$ being the size of the Roche lobe in units of the separation, as given, for example, in \citet{1983ApJ...268..368E}. In the case of stable mass transfer, the evolution proceeds until the envelope is lost, while the star, at least at the end of the evolution, is in thermal equilibrium \citep[e.g.][]{1995MNRAS.273..731R}. That means the star still obeys the core mass -- radius relation, and thus the period $P_m$ and mass of the remaining WD ($M_{WD1}$) after stable mass transfer can be obtained from the condition that the evolution ends with a Roche lobe that is equal to the radius of the giant with core mass equal to $M_{WD1}$. This leads to the well-known WD -- period relation that is observed in millisecond radio pulsars \citep[e.g.][]{1995MNRAS.273..731R,1999A&A...350..928T}. Because of the very similar dependence of $r_L(q)$ and $P$ on the mass of the companion, this relation is almost independent of $M_2$ and only depends on $M_{WD1}$ and metallicity. It can be approximated very well by \citep{2011ApJ...732...70L}
\begin{equation}
    P_{m, stable} = f \, 4.6 \, 10^6  \frac{M_{WD1}^9}{(1 + 25 \, M_{WD1}^{3.5} + 29 \, M_{WD1}^6)^{1.5}} \quad \text{d}
  ,\end{equation}
where the scale factor $f$ is 1.1 for solar metallicity and 0.8 for intermediate metallicity to match our radii (Eq.~(\ref{eq:R_Mc})). From the observed mass of the first formed WD ($M_{WD1}$), we can thus reconstruct the period after the first phase of mass transfer if that was stable. A similar approach is used by \citet{2021MNRAS.505.3514Z} to study sdB + helium WD binaries and  \citet{2024ApJ...977...24Z} for double WDs. \citet{2024ApJ...977...24Z} compared their semi-analytic model (using a slightly different core mass -- radius relation than ours) with detailed stellar-evolution calculations and found a small deviation at the lower end of the period or core mass range. In Sect.~\ref{sec:forward_model}, we show that this small difference does not affect our results.

Next, we can use the core mass -- radius relation again, but now using $M_{WD2}$ to determine the separation at the start of the second phase of mass transfer, because we assume  $M_{WD2}$ is equal to the core mass at the onset of mass transfer; that is, when the radius of the secondary was the size of its Roche lobe. This is possible because we assumed that the mass transfer is fast \citep[see][]{2000A&A...360.1011N}. It leads to
\begin{equation}
    a_{m, reconstruct} = \frac{R(M_{WD2})}{r_L(M_2/M_{WD1})}
,\end{equation}
from which we can obtain $P_{m, reconstruct}$ via Kepler's law. In this case, the reconstructed period does depend on $M_{WD2}$, metallicity, and the mass of the secondary $M_2$, because it is the Roche lobe of that star that we calculated. For each system, we simply calculated $P_{m, reconstruct}$ for the limiting masses of 1 and 2\msun from the separation given above.  We can now check if the reconstructed periods agree with those from stable mass transfer (i.e. $P_{m, reconstruct} =  P_{m, stable}$ for both metallicities). If so, the first phase of mass transfer could have been stable.

\subsection{Forward modelling of the expected masses}

In the second step, we calculated where in the $M_{WD2} - M_{WD1}$ plane we expect to find close double WDs for three different assumptions for the first phase of mass transfer: common envelope, stable mass transfer, or angular momentum balance using $\gamma$.

We start with a large number (30000) of uniformly and randomly selected values for the initial masses, $M_1$ and $M_2,$ plus the mass of the first formed WD, $M_{WD1,}$ to sample the possible progenitors, without taking into account the relative likelihood of the different progenitors.  We split the sample in two for the metallicity: 2/3 of the points have solar metallicity and 1/3 have intermediate metallicity (see Appendix~\ref{sec:Appendix_fits}). For each of the three different assumptions, we calculated the period after the first mass transfer ($P_m$). For the case of stable mass transfer we used the equations above to calculate $P_{m, stable}$, while for the common envelope and angular momentum balance we calculated $P_m$ by assuming the mass transfer was rapid, so it started when the primary already had a core mass equal to $M_{WD1}$ to calculate the initial period. We then used the standard equations to calculate $a_m$ \citep[e.g.][]{2005MNRAS.356..753N}:
\begin{equation}
    a_{m, CE} = a_i (1 - \Delta) \left(1 +
    \frac{2 q \Delta}{\alpha \lambda r_{\rm L}(q)}\right)^{-1}
,\end{equation}
\begin{equation}
    a_{m, AML} = a_i \left(\frac{1 - \gamma
      \frac{q \Delta}{1 + q}}{1 - \Delta}\right)^{2} \left(1 -
      \frac{q \Delta}{1 + q}\right),
\end{equation}
where $\Delta = (M_1-M_{WD1})/M_1$ is the envelope mass fraction, $\alpha \lambda$ is the common-envelope efficiency times a parameter describing the structure of the giant, and $\gamma$ is the fraction of specific angular momentum lost. From these, the periods after the mass transfer, $P_{m, CE}$ and $P_{m, AML,}$ are calculated.
The mass of the second WD is now given by the core mass of the giant of mass $M_2$ (we assume no mass is accreted, but that has little influence) that fills its Roche lobe in a binary with period $P_m$ and companion mass $M_{WD1}$.

\subsection{Reconstructing $\gamma$ values}

Finally, we also repeated the procedure of \citet{2000A&A...360.1011N,2005MNRAS.356..753N}, where the observed masses are used to reconstruct both the first and second phases of mass transfer, assuming the first was guided by the angular momentum balance and the second  was a common envelope. For each observed binary, the possible values of $\gamma$ that would give the right final masses can be calculated for a sample of progenitor masses, which we took as uniform between 1 and 2 \msun. We also sampled from the uncertainties in the observed masses of the two WDs. Each set of parameters gives rise to a different value of $\gamma$. The same procedure can be used to calculate the possible values of $\alpha \gamma$ in case the first phase of mass transfer was a common envelope as well.

\begin{table*}
    \caption{Overview of properties of the observed binaries, ordered by RA. Data were collected by \citet{2024MNRAS.532.2534M,2025MNRAS.540.1272M}.}
    \label{tab:observations}
    \small
    \begin{tabular}{lllllllllp{4cm}}
    id &name & $M_\mathrm{WD1}$ & $\sigma_\mathrm{WD1}$ & $M_\mathrm{WD2}$ & $\sigma_\mathrm{WD2}$ & $P$ &  $T_\mathrm{WD1}$ &  $T_\mathrm{WD2}$ &  Ref \\ 
     & & (\msun)  & (\msun)  & (\msun)  & (\msun) & (day) & (K) & (K) & \\
    \hline
    0 & WDJ000319.54+022623.28 & 0.38 & 0.025 & 0.47 & 0.02 & - & 7500 & 18200.0 & \citet{2025MNRAS.540.1272M}\\
    1 & WDJ002602.29-103751.86 & 0.42 & 0.02 & 0.47 & 0.02 & - & 5800 & 10700.0 & \citet{2025MNRAS.540.1272M}\\
    2 & WDJ005413.14+415613.73 & 0.45 & 0.04 & 0.43 & 0.04 & - & 7400 & 7700.0 & \citet{2025MNRAS.540.1272M} \\
    3 & WD0136+768 & 0.37 & - & 0.47 & - & 1.41 & 10500 & 18500.0 & \citet{2002MNRAS.332..745M}\\
    4 & WDJ  022558.21-692025.38 & 0.28 & 0.02 & 0.4 & 0.04 & 0.0328 & $\sim$14000 & 25330 & \citet{2023MNRAS.525.1814M} \\
    5 & SMSS  J033816.16-813929.9 & 0.38 & 0.05 & 0.23 & 0.015 & 0.0212 &  $\sim$10000 & 18100 & \citet{2021ApJ...918L..14K} \\
    6 & WD0455-295 & 0.44 & - & 0.4 & - & 0.3584 & $\sim$13000 & $\sim$16000 & \citet{1994ApJ...429..369W}\newline \citet{2020AA...638A.131N}
    \\
    7 & ZTF J0538+1953 & 0.45 & 0.05 & 0.32 & 0.03 & 0.0100 & 13000 & 26000 & \citet{2020ApJ...905...32B}
    \\ 
    8 & SDSS  J063449.92+380352.2 & 0.209 & 0.034 & 0.452 & 0.07 & 0.0184 & 10500 & 27300 &  \citet{2021ApJ...918L..14K} \\
    9 & ZTF  J0722-1839 & 0.38 & 0.04 & 0.33 & 0.03 & 0.0165 & 16800 & 19900 & \citet{2020ApJ...905...32B} \\
    10 & WD0957-666 & 0.32 & - & 0.37 & - & 0.0610 & 11000 & 30000 & \citet{1997MNRAS.288..538M} \newline \citet{2002MNRAS.332..745M}\\
    11 & CSS  41177 & 0.316 & 0.01 & 0.378 & 0.02 & 0.116 & 11678 & 24407 & \citet{2014MNRAS.438.3399B} \\
    12 & WD  1101+364  & 0.33 & - & 0.29 & - & 0.145 & - & - &\citet{1995MNRAS.275L...1M} \\
    13 & WDJ114446.16+364151.13 & 0.45 & 0.04 & 0.42 & 0.02 & - & 13000 & 13300.0 & \citet{2025MNRAS.540.1272M}\\ 
    14 & J1152+0248 & 0.325 & 0.013 & 0.362 & 0.014 & 0.0999 & 10400 & 20800 & \citet{2016MNRAS.458..845H} \newline  \citet{2020NatAs...4..690P}\\
    15 & WDJ135342.35+165651.75 & 0.43 & 0.05 & 0.47 & 0.04 & - & 7600 & 9600.0 & \citet{2025MNRAS.540.1272M} \\
    16 & WDJ141625.94+311600.55 & 0.42 & 0.02 & 0.47 & 0.03 & - & 12800 & 13300.0 & \citet{2025MNRAS.540.1272M} \\
    17 & WDJ141632.84+111003.85 & 0.42 & 0.02 & 0.47 & 0.03 & - & 7500 & 10500.0 & \citet{2025MNRAS.540.1272M} \\
    18 & WD  1447-190 & 0.33 & 0.09 & 0.41 & 0.1 & 1.790 & 5000 & 8000 &  \citet{2020MNRAS.493.2805K} \\
    19 & ZTF J1749+0924 & 0.4 & 0.07 & 0.28 & 0.05 & 0.0183 & 12000 & 20400 & \citet{2020ApJ...905...32B}\\ 
    20 & WDJ183442.33-170028.00 & 0.46 & 0.03 & 0.42 & 0.02 & - & 7000 & 8200.0 & \citet{2025MNRAS.540.1272M} \\
    21 & ZTF J1901+5309 & 0.36 & 0.05 & 0.36 & 0.05 & 0.0282 & 16500 & 26000 & \citet{2020ApJ...905...32B} \\
    22 & ZTF J1946+3203 & 0.272 & 0.046 & 0.307 & 0.097 & 0.0233 & 11500 & 28000 & \citet{2020ApJ...905...32B} \\
    23 & ZTF J2029+1534 & 0.32 & 0.04 & 0.3 & 0.04 & 0.0145 & 15300 & 18250 & \citet{2020ApJ...905...32B} \\
    24 & J2102-4145 & 0.314 & 0.01 & 0.375 & 0.01 & 0.100 & 12952 & 13688 & \citet{2023ApJ...950..141K}
    \\
    25 & WDJ211327.98+720814.03 & 0.38 & 0.02 & 0.42 & 0.02 & - & 7000 & 11100.0 & \citet{2025MNRAS.540.1272M} \\
    26 & WDJ212935.23+001332.26 & 0.44 & 0.02 & 0.44 & 0.04 & - & 7900 & 9200.0 & \citet{2025MNRAS.540.1272M} \\
    27 & ZTF J2243+5242 & 0.384 & 0.11 & 0.349 & 0.09 & 0.00611 & 16200 & 22200 & \citet{2020ApJ...905L...7B} \\ 
    28 & SDSS J232230.20+050942.06 & 0.24 & 0.06 & 0.27 & 0.06 & 0.0139 & 8000 & 19000  & \citet{2020ApJ...892L..35B} \\
    \end{tabular}
    \end{table*}

\section{Observations}\label{sec:observations}

As an observational sample, we used the list of double WD binaries collected by \citet{2024MNRAS.532.2534M,2025MNRAS.540.1272M}. We selected the binaries that have had both masses determined (either because they are double lined or because they are eclipsing) and for which both measured masses are below $0.49$\msun to ensure they are likely helium WDs. More massive stars can also form hybrid WDs (with C/O core and helium mantle) with masses below $0.49$\msun if they lose their envelope very early. However, given the narrow range of periods and the slope of the initial mass function (IMF), they should contribute at most a small fraction. The list consists of 18 objects from the literature and 11 recently discovered binaries  from the double-lined double-WD survey \citep[DBL,][]{2024MNRAS.532.2534M}. For the latter, only the two masses and the fact that they are in a (relatively) close binary are known; the orbital period is  not yet known. Because we only used the masses, this is not a problem for our method. We listed the parameters of the systems in Table~\ref{tab:observations}, using the updated masses from \citet{2025MNRAS.540.1272M}. For most systems, it is clear which of the two WDs is the younger one, given its higher temperature. However, for some systems the two temperatures are not so different, and WDs with different masses have different cooling curves. We used the results of \citet{2007MNRAS.382..779P} to assess if the formation order could have been different. Since more massive WDs cool slightly more slowly, we only needed to look at systems where the cooler WD is less massive or very similar in mass. System 2, WDJ0054, has two objects with very similar temperatures and mass, which could indicate that WD2 was formed first. A similar conclusion holds for systems 13 (WDJ1144), 16 (WDJ141625), and 24 (J2102). In system 12 (WD1101), the two WDs have very similar temperature/brightness, so we added this to the list of systems with uncertain formation orders. As we see below, this uncertainty does not influence our conclusions.

\section{Results}\label{sec:results}

\subsection{Reconstruction}

\begin{figure}
    \centering
    \includegraphics[width=\linewidth]{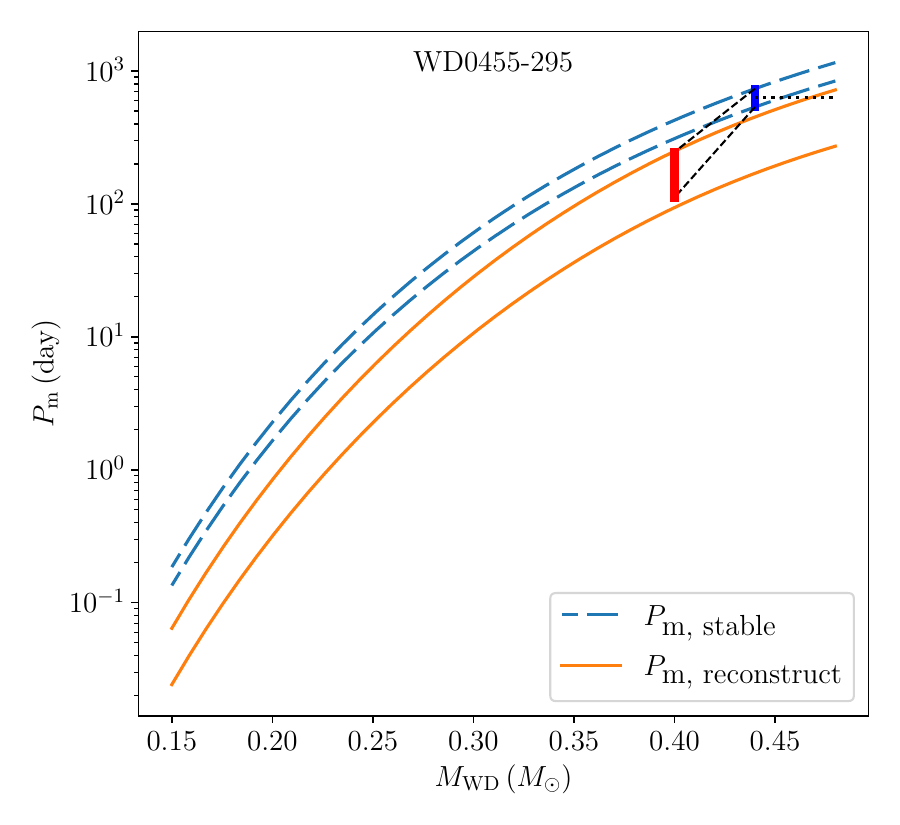}
    \caption{In orange: Reconstructed periods ($P_m$) as function of $M_{WD2}$. The top line shows $M_2 =1$\msun and solar metallicity, and the bottom line shows  $M_2 =2$\msun and intermediate metallicity. The vertical orange bar indicates the possible values of $P_m$ for  WD0455-295.(source id 6) based on the mass of the youngest WD ($M_\mathrm{WD2} = 0.4$\msun). The dashed blue line shows expected periods after stable mass transfer as function of $M_{WD1}$. The position of  WD0455-295 is indicated by the vertical blue bar at the position of its mass of the first formed WD ($M_\mathrm{WD1} = 0.44$\msun). The two periods are not consistent, as shown by the dotted horizontal line at $P_{m, stable}$ .}
    \label{fig:Pm_single}
\end{figure}

\begin{figure}
    \centering
    \includegraphics[width=\linewidth]{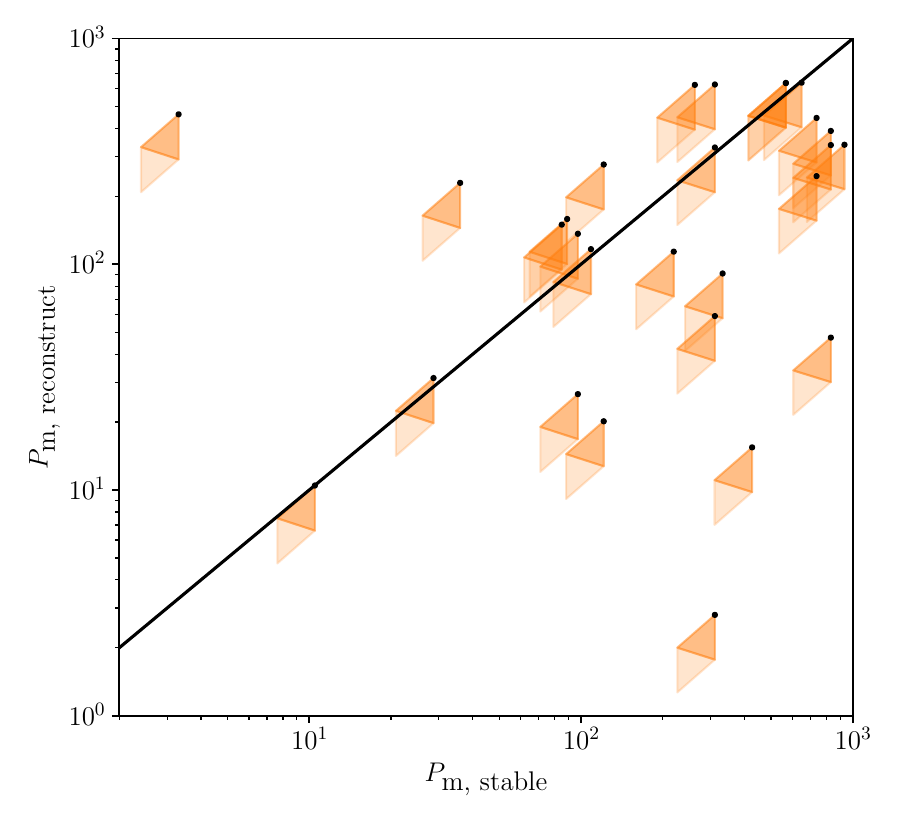}
    \caption{Reconstructed intermediate period based on reconstructing the second phase of mass transfer  (y-axis) versus expected intermediate period based on the assumption that the first phase of mass transfer was stable (x-axis). The diamond shapes indicate the possible values for the observed systems for different assumptions. Lowering the metallicity lowers both periods, while increasing $M_2$ lowers only the values on the y-axis. 
 Top right of each diamond thus is for solar metallicity and $M_2 = 1$\msun and bottom left is for intermediate metallicity and $M_2 = 2$\msun. The shading indicates that, very broadly, the upper parts are more likely scenarios (see text). If all systems were formed through stable mass transfer, all shapes should intersect with the diagonal. Systems 1, 16, and 17 have identical masses and thus fall on top of each other.}
    \label{fig:Pm_reconstruct}
\end{figure}

\begin{figure*}
    \sidecaption
    \centering
    \includegraphics[width=11cm]{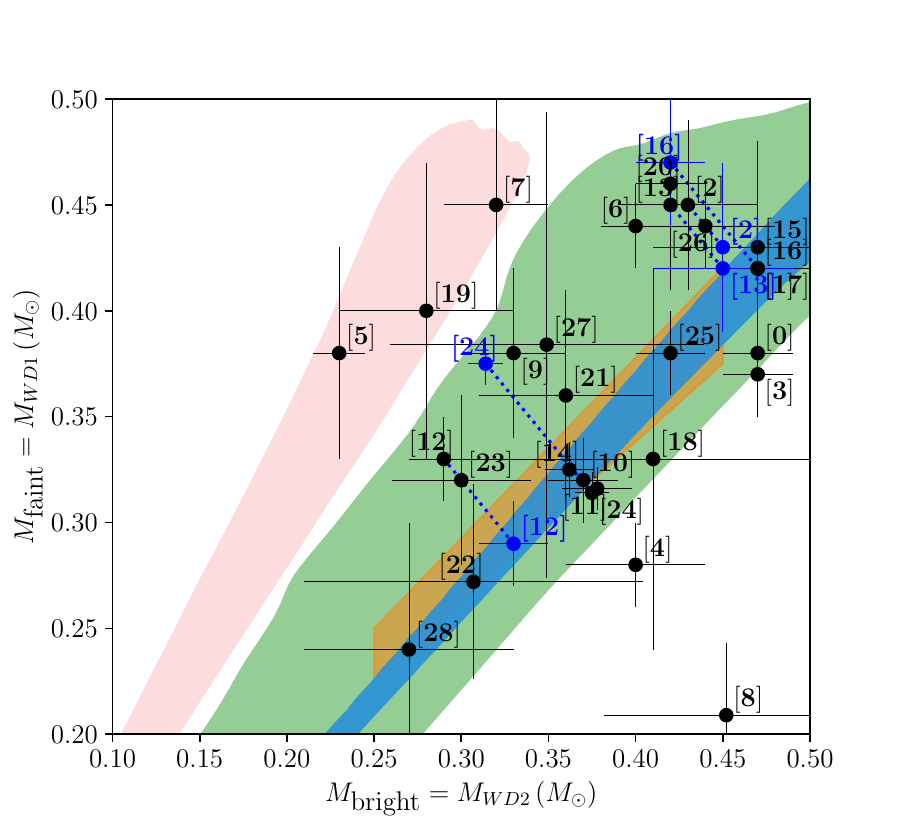}
    \caption{Predicted mass distribution based on different assumptions for the first phase of mass transfer. In dark blue, we show the predicted masses in the case that the first phase of mass transfer was stable. The orange area underneath covers the range for stable mass transfer as found by \citet{2024ApJ...977...24Z}. The results for the case where the first phase of mass transfer was a common envelope with $\alpha \lambda = 2$ is shown in light salmon on the far left, and those for the case where the first phase was described by the angular momentum balance with $\gamma = 1.6$ are shown in green. The data points and their uncertainties are plotted in black, with the labels indicating the ID of the source in Table~\ref{tab:observations}. The four data points for which the order of formation is not completely clear are plotted in blue and connected with dotted lines with the masses reversed (see text). Systems 1, 16, and 17 have identical masses and thus fall on top of each other.}
    \label{fig:M1m2}
\end{figure*} 

As an illustration of the reconstruction process, we show, in Fig.~\ref{fig:Pm_single}, the example for WD0455-295 (source ID 6 in Table~\ref{tab:observations}). The dashed blue lines show $P_{m, stable}$ for solar- (top) and low-metallicity (bottom). At the observed mass of the first formed, currently cooler WD ($M_{WD1} = 0.44$ \msun), the range of $P_{m, stable}$ is indicated by the blue bar (535-735 d). The solid orange lines represent the reconstructed periods, $P_{m, reconstruct}$, as a function of WD mass (i.e. the period in which the secondary fills its Roche lobe with a core mass equal to the x-coordinate). The upper line is for a secondary mass of 1 
\msun at solar metallicity, and the bottom line is for 2 \msun star at intermediate metallicity. The reconstructed period range for WD0455-295 is 110-245 d and is shown as the orange bar at the mass of the second formed, currently hotter WD ($M_{WD2} = 0.4$ \msun). The horizontal dotted line shows the expected period at Roche-lobe overflow if the first mass transfer had been stable. It is clear that that is not correct. 

We then repeated the procedure for all sources, but rather than plotting a figure for each source, we put them all in one plot. In Fig.~\ref{fig:Pm_reconstruct}, for each source we show a diamond shape with  the range of $P_{m, stable}$ on the x-axis (equivalent to the blue bar in Fig.~\ref{fig:Pm_single}) and the range of $P_{m, reconstruct}$ on the y-axis (equivalent to the orange bar in Fig.~\ref{fig:Pm_single}), where we take into account that the metallicities should be the same for both first and second mass transfer. The top right corner of the diamond (with the black dot) is for Solar metallicity and a secondary mass of 1 \msun, which we consider the more likely progenitor range. Lowering the metallicity lowers both periods (top left corner), while increasing $M_2$ lowers only the intermediate period reconstructed from the second phase (bottom right corner). The bottom left corner of the diamond is thus for intermediate metallicity and a 2 
\msun secondary, which we consider less likely (since there are fewer more massive stars due to the IMF and most of the stars in the Milky Way have Solar metallicity). We shade the top of the diamonds darker to indicate this higher likelihood.

If all sources had been formed through stable mass transfer, we would expect all the diamonds to intersect with the diagonal $P_{m, reconstruct} =  P_{m, stable}$. This is clearly not the case, although some sources are consistent with the first phase being stable. In this analysis, we did not take the uncertainties in the masses of the WDs into account; this is done in the next section.

\subsection{Forward modelling}\label{sec:forward_model}

We now look at the position in the $M_{WD2} - M_{WD1}$ parameter space where the close double WDs would end up if the first phase of mass transfer were stable, a common envelope, or governed by the angular-momentum balance $\gamma$. For the common-envelope efficiency, we used $\alpha \lambda = 2$; and for the angular-momentum balance, we used $\gamma = 1.6$ (see Sect.~\ref{sec:reconstruct_gamma}). In Fig.~\ref{fig:M1m2}, we show the resulting distribution plots for the different models, overplotted with the data points, where we can now take the uncertainties of the data into account. We stress again that our sampling of the progenitor parameter range does not take a relative likelihood such as the IMF into account. Similarly, we did not consider selection effects on the parameters of the observed binaries. The question here is simply whether the different models can reach the combination of masses of the observed systems or not. 

Regarding the assumption that the first phase was a common envelope (light salmon coloured distribution), we confirm the earlier findings that this is a very poor fit to the data. In these models, the fainter WD is always significantly more massive than the brighter WD, while in the observations this is very rare. Indeed, only SMSS J0338 (ID 5), ZTF J0538 (ID 7), and ZTF J1749 (ID 19) have masses that are compatible with a two-common-envelope evolution. 

Conversely, the model in which the first phase of mass transfer was stable (dark blue distribution) predicts a narrow range of masses in which the bright WD is somewhat more massive than the faint WD. We note that this region is set by the core mass -- radius relation and hardly depends on the mass transfer physics such as how conservative the mass transfer is \citep[see also][]{2024ApJ...977...24Z}. Although about half of the systems have masses compatible with this prediction, it is clear that overall the model is not a good match with the full set of observations; this is consistent with what we found above. The detailed models of \citet{2024ApJ...977...24Z} find slightly lower mass ratios at low WD masses. Their mass range is shown as the orange area underneath the blue one and does not change; rather, it supports our conclusion. The parameter range covered by the model using the $\gamma$ prescription (green distribution) fits better with the data. One system, SDSS J0634, cannot easily be explained by any of the models as its brighter WD is much more massive than the fainter one.

We used ConTEST \citep{2023MNRAS.524.1061S} to carry out a formal consistency test of the data with respect to the models, despite the fact that we did not take the initial distributions of the progenitor parameters into account. The detailed results are given in Appendix \ref{sec:Appendix_ConTEST} and show that the stable mass transfer and double CE model are firmly rejected (p-value 0.0), while the model using the $\gamma$ prescription is formally consistent with a p value of 0.13 if we exclude the data points that are consistent with two common envelopes (IDs 5, 7, and 19). We note that we had to remove all WD2 masses below $0.27$\msun in the models since there is no support for these values in the data, which may be due to selection effects.

To assess the impact of a possible reversal of the masses in the systems in which the two WDs have very similar temperatures and ages, we plot five systems (systems 2, 12, 13, 16, and 24 in Table~\ref{tab:observations}), with the masses swapped in blue in Fig.~\ref{fig:M1m2}. The conclusions do not change. In fact, the situation worsens.

\subsection{Reconstructing $\gamma$ values}\label{sec:reconstruct_gamma}

\begin{figure}
    \centering
    \includegraphics[width=\linewidth]{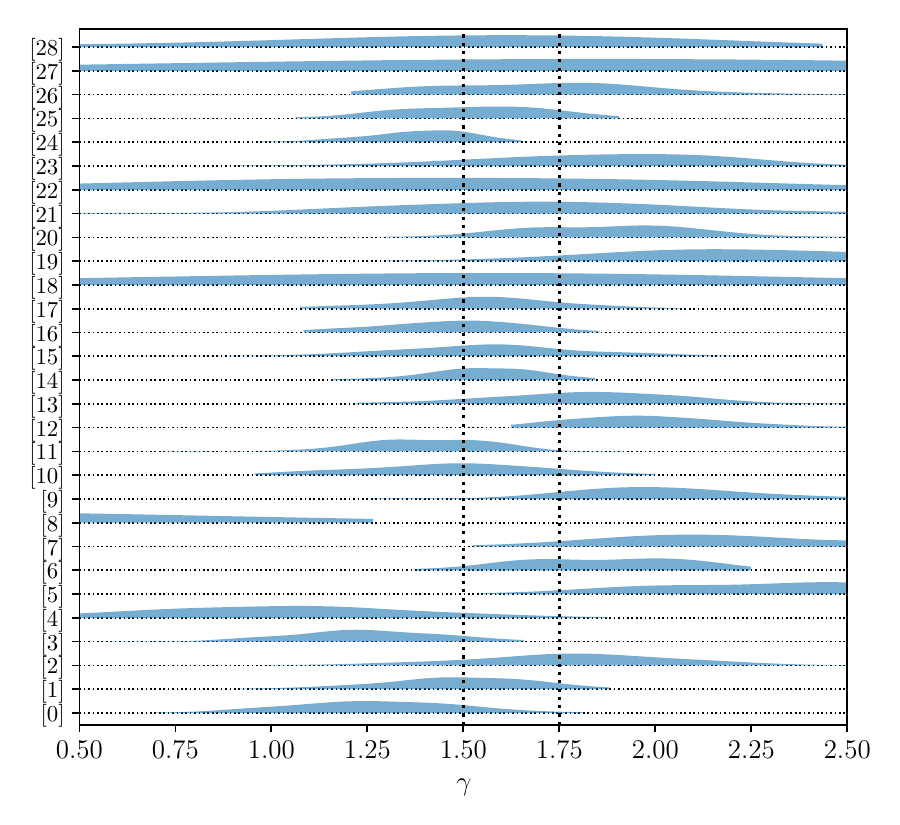}
    \caption{Kernel density estimate of reconstructed $\gamma$ values for each of the systems in Table~\ref{tab:observations} for a set of different values of initial masses and metallicity as well as different WD masses following the uncertainties in the derived masses. The KDE combines values for specific progenitor masses and bracket the range of possible $\gamma$ values for the range of progenitor masses and the uncertainties in the WD masses. }
    \label{fig:gamma_values}
\end{figure}

Following up on the previous finding that most systems can be explained by a model using the $\gamma$ prescription, we reconstructed the values of $\gamma$ for each of the observed systems, taking the observed masses into account, sampling the uncertainties assuming a normal distribution and flat distribution of initial masses between 1 and 2 \msun. The results are plotted in Fig.~\ref{fig:gamma_values}, where we show a kernel density estimate of  120 random selections from the parameter distribution (80 for solar and 40 for intermediate metallicity). It confirms that all systems apart from system 8 are compatible with a value of $\gamma$ between 1.5 and 1.75. We note that the simulation samples the full uncertainty, so it contains systems outside the 1$\sigma$ region. That is why source 5, for example, falls far from the green area in Fig.~\ref{fig:M1m2}, but still has solutions with $\gamma =1.75$.

The possible solutions of $\alpha \lambda$ assuming the first phase of mass transfer was also a common envelope are shown in Appendix~\ref{sec:Appendix_al}. For 13 systems, there are no solutions with positive values, while for another eight there are only solutions  with $\alpha \lambda > 2$. This confirms the conclusion that two common-envelope phases are a poor fit to the observations.

\section{Discussion and conclusions}\label{sec:conclusion}

In our analysis, we made a number of assumptions that could affect the results. First, we used an analytic core mass -- radius relation, rather than performing detailed stellar evolution calculations. Previous work has shown that the final outcome of stable mass transfer typically follows the core mass -- radius relation tightly \citep[e.g.][]{1995MNRAS.273..731R,1999A&A...350..928T,2011ApJ...732...70L}. The recent detailed study of \citet{2024ApJ...977...24Z} is exactly targeted at the evolution we studied. They do find that the mass ratio of the final double-helium WD is somewhat lower at a low WD mass compared to their semi-analytic model ($q = 1-1.1$ instead of $q = 1.2-1.3$). They conclude that this is due to a small amount of additional hydrogen mass on top of the core (making the first WD more massive) as well as small deviations of the core mass -- radius relation for stars that started mass transfer before or at the base of the red giant branch. Interestingly, our semi-analytical model, using a slightly different core mass -- radius relation, gives mass ratios of $q = 1.1-1.2$ at low WD masses; that is, between their semi-analytic and detailed models. In Fig.~\ref{fig:M1m2}, we indicate the region of the detailed models of \citet{2024ApJ...977...24Z} in orange, showing the small offset but supporting our conclusion. 

\citet{2023MNRAS.525.2605G} investigated the effect of stellar wind and found that normal winds do not change the core mass -- period relation significantly, but strong, tidally enhanced stellar winds can lead to significantly wider periods, as the donor can lose Roche-lobe contact. In principle, this could widen the parameter range from stable mass transfer, but only towards more unequal masses and not to systems with equal masses or reversed mass ratios. 

More generally, we assumed that the period of the binary does not change between the end of the first phase of mass transfer and the onset of the second phase of mass transfer. Apart from (strong) winds, this could also be influenced by tides if there is significant transfer of orbital angular momentum to the spin of the secondary. This would shrink the orbit and bring the mass of the second formed WD closer to the mass of the first formed one, but it would preserve the tight relation. 

Additionally, we assumed that all observed double-WD systems with masses below $0.49$\msun originate from mass transfer of stars with initial masses between 1 and 2 \msun that begin mass transfer after the main sequence (case B). The relations we derive do not hold for more massive stars (that do not have degenerate helium cores) and for mass transfer that starts on the main sequence. A full population model will thus add different formation channels for double WD that add systems to the mass range we considered and will likely populate other regions. The IMF and initial period distribution suggest that the majority of systems follow the evolution outlined here. However, there seems to be a cluster of systems with relatively high masses (total mass around 0.9\msun), perhaps indicating a contribution from more massive progenitors. Furthermore, we only used two metallicity models. Using a broader range of metallicities would widen the range of periods that are possible after stable mass transfer. However, our metallicities cover the majority of the star formation that has happened in our Milky Way, so we do not believe that this will change the conclusions of our work. Finally, we assumed that the mass of the core does not grow during the second phase of mass transfer, which we believe is well justified by the expected short duration of the common-envelope phase.

\begin{figure}
    \centering
    \includegraphics[width=\linewidth]{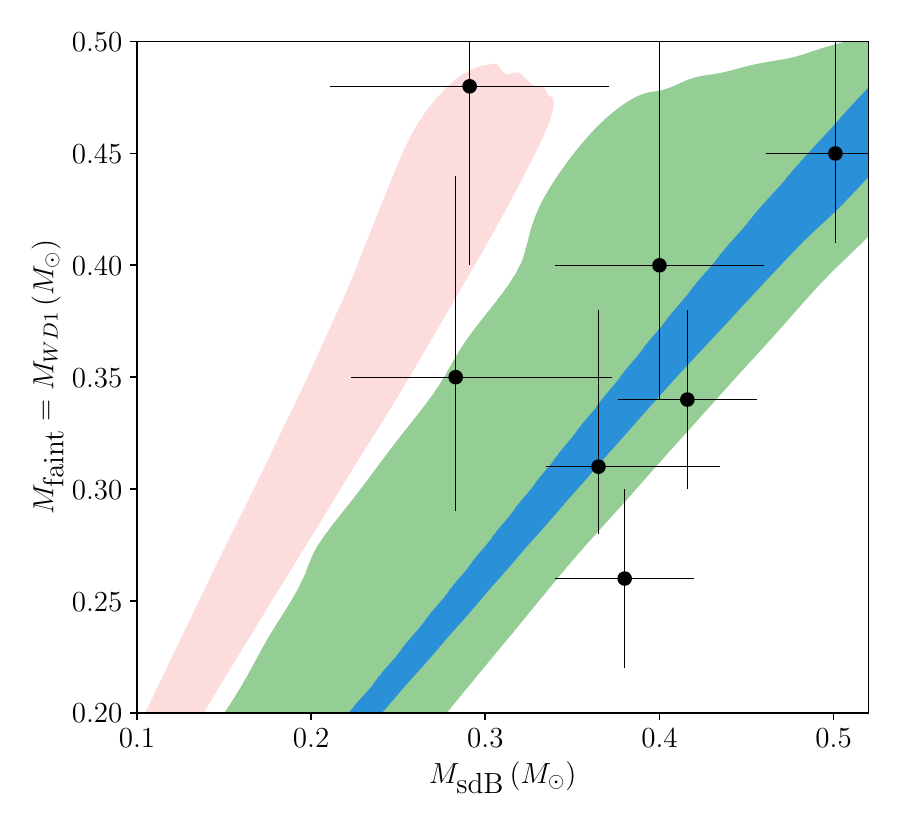}
    \caption{Same as Fig~\ref{fig:M1m2}, but with the masses of the sdB + WD binaries overplotted. }
    \label{fig:M1m2_sdB}
\end{figure}

Our results seem at odds with the conclusions of \citet{2024ApJ...961..202G} and \citet{2024ApJ...977...24Z} that the formation of double WD and sdB + WD binaries points at a stable initial mass transfer. They also used the core mass -- radius relation to infer the periods before the second phase (which is assumed to be a common envelope) and selected the progenitor from a set of models that follow the core mass -- radius relation. \citet{2024ApJ...977...24Z} compared their resulting mass ratios qualitatively with those measured in double WD, and for our systems 10 and 11 they carried out a more detailed study (mainly to determine the efficiency of the final common envelope). These two are indeed consistent with stable mass transfer. To study the sdB + WD binaries, we plot the masses of the sdB and WD in the systems, mainly taken from \citet{2022A&A...666A.182S,2023A&A...673A..90S}, in Fig.~\ref{fig:M1m2_sdB}. \citet{2024ApJ...961..202G} assumed the two objects with low sdB mass are formed from stars more massive than 2\msun that have non-degenerate cores and that the others could indeed be consistent with the assumption that the first phase was stable, but sometimes there is a lack of agreement (e.g. they find periods that differ from the predicted period by up to 60\%). In light of our conclusion above that the first phase of mass transfer cannot always have been stable for double WDs that very much follow the same evolution, we believe the sdB + WD binaries likely also did not all go through a stable first phase. Indeed, the distribution of data points in Fig.~\ref{fig:M1m2_sdB} is similar to that in Fig.~\ref{fig:M1m2}. \citet{2021MNRAS.505.3514Z} used a very similar method to ours to construct the final helium WD mass - period relation for helium WD + sdB binaries. They concluded that stable mass transfer is in broad agreement with the observations. However, they made the comparison more complicated by including the final common-envelope phase, and their model proves not to fit the data very well.

\citet{2023A&A...669A..82L} simulated the Galactic population of double WDs using stability criteria that favour stable mass transfer and found that the results provide a smaller population that fits the observational constraints better. Surprisingly, the resulting mass-ratio distribution of the double WD (their Figure 7) has a strong peak at equal masses, but not at $q > 1$, in contrast to our results. This may be related to the treatment of stable mass transfer.

\begin{figure}
    \centering
    \includegraphics[width=\linewidth]{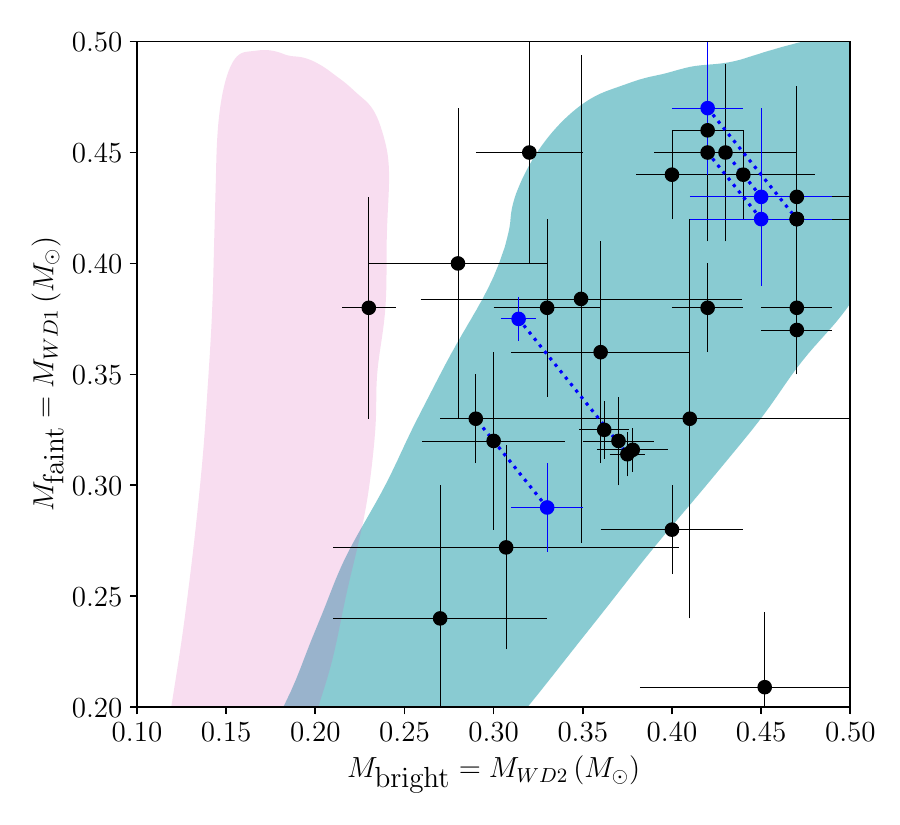}
    \caption{Predicted mass distribution based on the SCATTER formalism. The light pink distribution is constructed using the parameters found in \citet{2023ApJ...944...87D}. The fit to the data is even worse than for the standard common-envelope case. For the alternative parameter $\eta = 2$, the model can produce a first phase of mass transfer that is similar to that using the $\gamma$ prescription (darker cyan). }
    \label{fig:M1m2_SCATTER}
\end{figure}

\citet{2023ApJ...944...87D} suggested an approach to the common envelope that is also based on angular momentum, but with a more physical interpretation. The assumption is that the angular momentum that is taken with the material lost from the binary is the sum of what is provided by the core of the giant and what is provided by the companion. The model depends on a parameter $\eta$ that multiplies the specific angular momentum that is transferred to the envelope (for an assumed value of the dimensionality parameter $\delta = 3$). Based on known post-common-envelope systems, \citet{2023ApJ...944...87D} found that $\eta$ is large and a function of the ratio of the total mass and envelope mass, with systems with larger ratios  having larger $\eta$. We investigated the results if the first phase of mass transfer could be described by their fit to $\eta = 4 (M_{total}/M_{envelope})$ and show the resulting mass distribution in Fig.~\ref{fig:M1m2_SCATTER} as the light pink distribution. Since this relation was fitted to systems that underwent clear shrinkage of their orbit, it is perhaps not surprising that the results with such high $\eta$ values look very much like the case where the first phase of mass transfer was described by the standard common-envelope equations. We do find that if the first phase of mass transfer was described by this formalism but with a constant value $\eta = 2$, the resulting masses (dark cyan distribution in Fig.~\ref{fig:M1m2_SCATTER}) are similar to those found for the $\gamma$ formalism and even somewhat more extended, potentially covering more of the observed systems.

In summary, we used two methods to investigate if the first phase of mass transfer leading to double-helium WD binaries was stable and conclude that this is not generally the case. Where stable mass transfer is expected to lead to a well-defined period after the first mass transfer and a strong correlation between the masses of the final two WDs, the observations suggest a very broad range of periods after the first phase of mass transfer. Despite its lack of a clear physical interpretation, the $\gamma$ prescription still gives the best match of the models to the observations. 

The mismatch between the periods after the first mass transfer and models assuming stable mass transfer may very well be related to the finding that many of the observed binaries that are in the phase (just) after the first mass transfer have eccentric orbits, while the previous mass transfer is expected to lead to circular orbits for all models we considered here \citep[e.g.][]{2008A&A...480..797B,2025ApJ...984..137D}. Several solutions have been investigated \citep[e.g.][]{2008A&A...480..797B,2013A&A...551A..50D,2015A&A...579A..49V,2019A&A...629A.103S,2020A&A...639A..24E,2020A&A...642A.234O}, many of which have problems and several rely on the donor having been an AGB star undergoing strong stellar-wind mass loss. However, the fact that descendants of RGB donors can also have eccentric orbits and, as we show here, do not seem to follow stable mass transfer, suggests that the problem is not unique to AGB donors.  The solution to that conundrum can perhaps also explain the masses of the observed double-WD population.

\begin{acknowledgements}
We thank the referee for constructive comments that inproved the article. This work was supported by the Netherlands Science Foundation NWO and the Radboud Excellence Initiative. 
    
\end{acknowledgements}

\bibliographystyle{aa}
\bibliography{low_mass_stability}

\appendix

\section{Period/radius - core mass relations}\label{sec:Appendix_fits}

\begin{figure}[h]
    \centering
    \includegraphics[width=\linewidth]{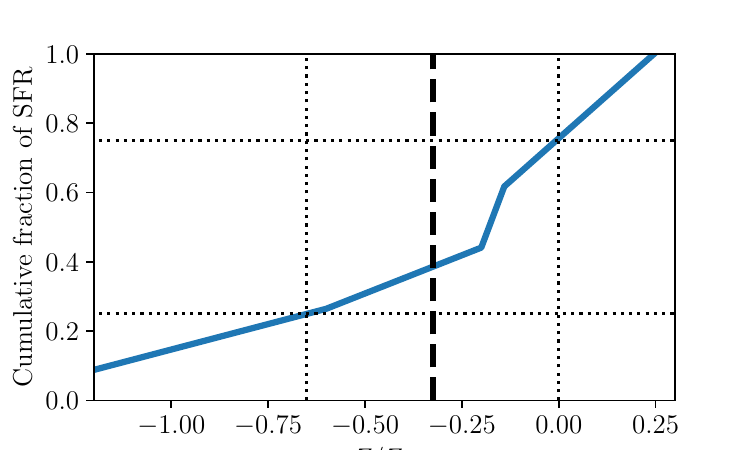}
     \caption{Cumulative fraction of the star formation in the Milky Way as function of metallicity based on  \citet{2015A&A...578A..87S}. In their calibration, models with metallicity $Z=0.02$ ($[Z/Z_\odot]= 0$) and $Z=0.0045$ ($[Z/Z_\odot] = -0.65$) cover the 25\% percentiles. A two-metallicity model has about 1/3 of the SFR in the lower metallicity bin (left of the dashed line) and about 2/3 in the higher metallicity bin.}
      \label{fig:SFR_cumul}
\end{figure}

Several studies, in particular into the formation of millisecond radio pulsars with WD companions have determined the period -- core mass/WD mass relation that is expected after a phase of stable mass transfer. \citet{1995MNRAS.273..731R} gave a simple relation, but \citet{1999A&A...350..928T} and \citet{2011ApJ...732...70L} found that at low mass that relation overestimates the periods. In addition, the radius of the giants and thus the period -- core mass relation depends on metallicity. \citet{1995MNRAS.273..731R} and \citet{1999A&A...350..928T} find that population II stars, with assumed metallicity of 0.001, i.e about 1/20 of Solar, have about 40\% smaller radii. However, such low metallicity is very rare in the Milky Way. In Fig.~\ref{fig:SFR_cumul} we plot the cumulative distribution of the metallicity of the total amount of star formation in the Milky Way, based on the derived star formation rate and metallicity by \citet{2015A&A...578A..87S}. It can be seen that there is a similar amount of star formation with metallicities higher than Solar as there is with metallicities below  $Z=0.0045$, which on a logarithmic scale lies between the two metallicities used in the models by \citet{1999A&A...350..928T}. We therefore use two metallicities in this study: Solar and intermediate in which the radii are 20\% smaller than the Solar values. The latter is similar to "Pop I+II" in \citet{1999A&A...350..928T}, For the population studies we also use these two metallicities as metallicity bins. The lower metallicity covers about 1/3 of the total SFR (see Fig.~\ref{fig:SFR_cumul}). 

We adapted the curve of \citet{2011ApJ...732...70L} that is given only for solar metallicity  and for relatively low mass to match the ones of \citet{1995MNRAS.273..731R,1999A&A...350..928T} at higher mass. For this we multiply their relation with a factor 1.1 for solar metallicity and a factor 0.8 for intermediate metallicity. We note that \citet{2023MNRAS.525.2605G} find somewhat larger periods. 

\begin{figure}
    \centering
    \includegraphics[width=\linewidth]{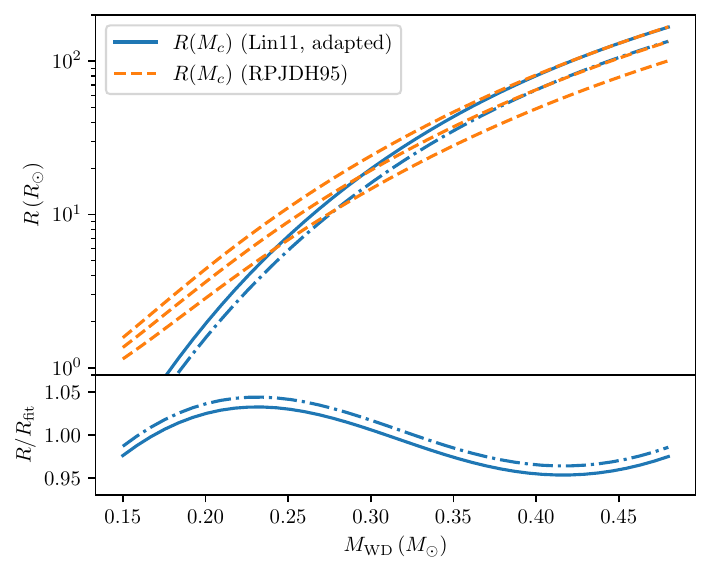}
     \caption{Models from \citet[their eq. (5)]{1995MNRAS.273..731R}, orange dashed, for $R0$ equal to 5500, 4400 and 3300 (top to bottom) which at high core mass are very similar to the Pop I, Pop I+II and Pop II curves of  \citet{1999A&A...350..928T}. Our version of the \citet{2011ApJ...732...70L}  curves that are adapted to match the other top two curves are shown in blue. The bottom panel shows the residuals between the adapted curves and our simple fit that show that the fit is accurate to within 5 per cent.}
      \label{fig:Rfit}
\end{figure}

We use the adapted period -- core mass relation of \citet{2011ApJ...732...70L} to reconstruct the core mass -- radius relation. This is shown in Fig.~\ref{fig:Rfit}, compared to the core-mass radius relation of \citet{1995MNRAS.273..731R}. We fit the relation with a simple function:
\begin{equation}
    R = R_* \frac{M_c^6}{(1 + 45 M_c^5)} R_\odot
\end{equation}
with $R_* = 3 \times 10^4$ for solar  nd $2.4 \times 10^4 $ for intermediate metallicity. The fit is good to within a few per cent (Fig.~\ref{fig:Rfit}).

\section{Reconstructed $\alpha \lambda$ values}\label{sec:Appendix_al}

\begin{figure}
    \centering
    \includegraphics[width=\linewidth]{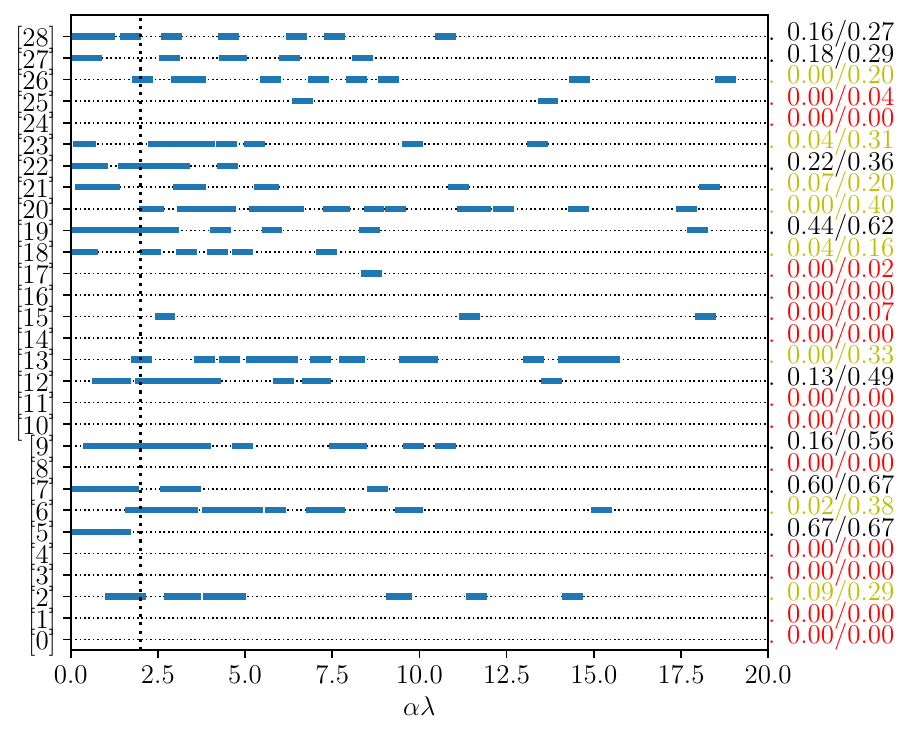}
    \caption{Reconstructed $\alpha \lambda$ values for each of the systems in Table~\ref{tab:observations} assuming that the first phase of mass transfer was a common envelope, for a set of different values of initial masses and metallicity as well as different WD masses following the uncertainties in the derived masses. The numbers  on the right show the fraction of solutions that falls in the 0-2 range and the 0-20 range respectively. Systems that have less than 10\% solutions in the 0-20 range are coloured red, ones that have less than 10\% solutions in the 0-2 range yellow, the others black. }
    \label{fig:al_values}
\end{figure}

\renewcommand{\thefigure}{C.1}
\begin{figure*}[h]
    \includegraphics[width=12cm]{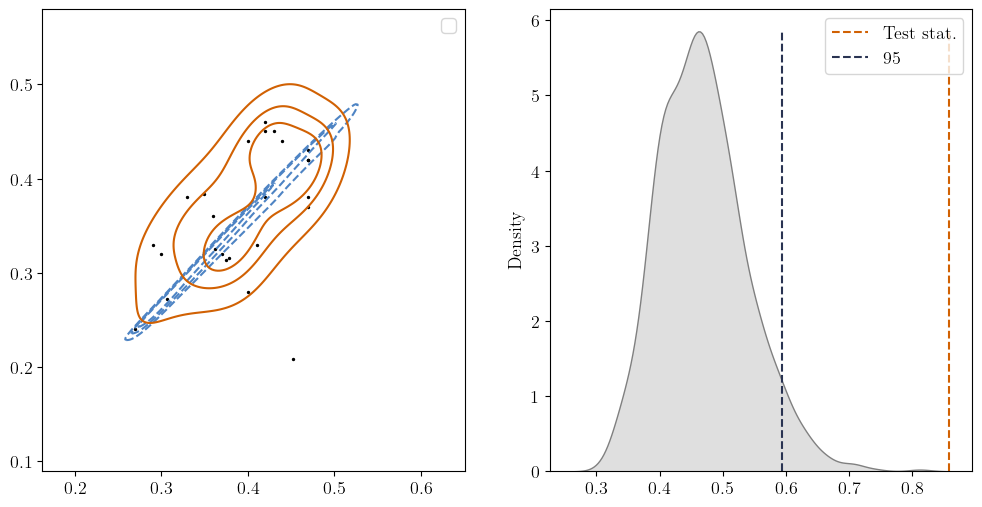}\\
    \sidecaption
    \includegraphics[width=12cm]{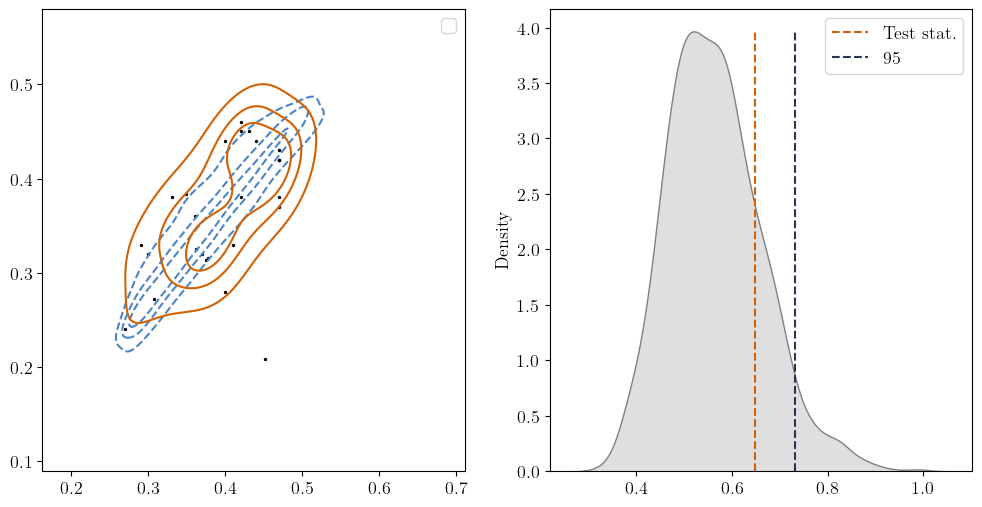}
    \caption{Left panels: densities determined by ConTEST for the data (orange, solid) and model (dashed, blue). Right panels: distribution of distances between simulated data sets and model. The 95\% limit is shown as black dashed line, the distance of the actual observations is shown as orange dashed line. Top panels are for the model assuming the first mass transfer is stable. The model is rejected as the distance is larger than any of the simulated data sets (p-value is 0.0). Bottom for the model assuming the first phase of mass transfer is described by the $\gamma$-prescription. In this case the model is consistent, with p-value of 0.13. }
    \label{fig:ConTEST}
\end{figure*}

For completeness, we also reconstruct the necessary values for $\alpha \lambda$ for the assumption that the first phase of mass transfer was a standard common envelope. In Fig.~\ref{fig:al_values} we plot the solutions where we sample 45 points using the same procedure as in Sect.~\ref{sec:reconstruct_gamma}. The numbers show the fraction of solutions for $\alpha \lambda$ in the ranges 0-2 and 0-20 respectively. The 13 red systems do not find more than 10\% solutions in the 0-20 range, and we conclude they cannot be explained by a common envelope, the 8 yellow ones not in the 0-2 range and here common envelope is very unlikely. This confirms the conclusion that two common-envelope phases is a very poor model for the full set of observed systems.

\section{Consistency tests of the different models}\label{sec:Appendix_ConTEST}

In order to more formally compare the models with the data we use the ConTEST package developed by \citet{2023MNRAS.524.1061S}. This includes several tests, including \verb|contest_den| that is aimed at comparing data points with a 2D density distribution. We use this on our models and data, with a sampling value $K=1000$. In Fig.~\ref{fig:ConTEST} we show in the top panel the ConTEST for stable mass transfer. The model is rejected with a p-value of 0.0. For this test we even removed the three systems that are compatible with two common-envelope phases and removed from the model all data with $M_{WD2} < 0.27$, since the test doesn't find any density there based on the observations and these could be missing due to selection effects. The same test on the double CE model (now including all data points) also finds a p-value of 0.0. The model using the $\gamma$-prescription is formally compatible with the data if we again remove the 3 data points that are compatible with double common-envelope evolution and remove from the model all data with $M_{WD2} < 0.27$, with p-value 0.13. Note that we don't remove the system with id 8, that by ConTEST is identified as an outlier.

\end{document}